\documentclass[conference]{IEEEtran}
\usepackage[pdftex]{graphicx}
\usepackage{graphicx}
\usepackage{balance}
\usepackage{comment}
\usepackage{algorithm}
\usepackage[noend]{algpseudocode}
\usepackage[dvipsnames]{xcolor}
\usepackage{cite}
\usepackage{amsmath}
\usepackage{amssymb}
\usepackage{booktabs}
\usepackage{lipsum}
\usepackage[capitalise]{cleveref}
\usepackage{subcaption}
\newcommand{\url}{\texttt}
\crefname{equation}{}{}

\usepackage{acronym}
\acrodef{AoA}{Angle of Arrival}
\acrodef{RSSI}{Received Signal Strength Indicator}
\acrodef{SVM}{Support Vector Machine}
\acrodef{KNN}{K-Nearest Neighbors}
\acrodef{RF}{Random Forest}
\acrodef{SF}{superframe}
\acrodef{HMM}{Hidden Markov Model}
\acrodef{MLC}{machine learning classifier}
\acrodef{BLE}{Bluetooth\textsuperscript{\textregistered} Low Energy}
\acrodef{BT}{Bluetooth\textsuperscript{\textregistered} Classic}
\acrodef{ephesos}[EPhESOS]{Energy and Power Efficient Synchronous Sensor
Network}
\acrodef{PHY}[PHY]{physical}
\acrodef{BS}{base station}
\acrodef{MAC}{media access control}
\acrodef{TDMA}{time division multiple access}
\acrodef{ToA}{Time of Arrival}
\acrodef{TDoA}{Time Difference of Arrival}
\acrodef{TSN}{Time-Sensitive Networking}
\acrodef{SN}{sniffer node}
\acrodef{PS}{photo sensor}
\acrodef{MN}{measurement node} 
\acrodef{SAL-RB-Dataset}{SAL Autarkic Localization RSSI BLE Dataset}
\acrodef{RBF}{radial basis function}
\acrodef{IWSN}{industrial wireless sensor network}
\acrodef{WSN}{wireless sensor network}
\acrodef{GPS}{Global Positioning System}
\acrodef{UWB}{ultra-wideband}
\acrodef{WLAN}{Wireless Local Area Network}
\acrodef{I40}{Industry 4.0}
\acrodef{IoT}{Internet of Things}

\acrodef{LLR}{log likelihood ratio}
\acrodef{MWIS}{Maximum Weighted Independent Set}
\acrodef{ILP}{Integer Linear Programming}
\acrodef{cca}[CCA]{Clear Channel Assessment}
\acrodef{tpr}[TPR]{true positive rate}
\acrodef{tnr}[TNR]{true negative rate}
\acrodef{fpr}[FPR]{false positive rate}
\acrodef{fnr}[FNR]{false negative rate}
\acrodef{RMSE}[RMSE]{root mean squared error}

\acrodef{MHT}{Multi Hypothesis Tracking}
\acrodef{ISM}{industrial, scientific and medical}

\acrodef{MAM}{multiply, add, and modulo}
\acrodef{csa1}[CSA\#1]{Channel Selection Algorithm \#1}
\acrodef{csa2}[CSA\#2]{Channel Selection Algorithm \#2}
\acrodef{GFSK}{Gaussian Frequency Shift Keying}
\acrodef{FHSS}{frequency hopping spread spectrum}
\acrodef{sdr}[SDR]{Software Defined Radio}
\acrodef{wpan}[WPAN]{Wireless Personal Area Network}

\acrodef{ecdf}[eCDF]{empirical Cumulative Distribution Function}
\acrodef{eccdf}[eCCDF]{empirical Complementary Cumulative Distribution Function}

\usepackage{tikz}
\usetikzlibrary{arrows,matrix,positioning}
\tikzstyle{int}=[draw, minimum size=2em] 
\tikzstyle{init} = [pin edge={to-,thin,black}]

\IEEEoverridecommandlockouts

\begin{document}

\title{
Predicting the Channel Access of \\ Bluetooth Low Energy
}

 \author{\IEEEauthorblockN{Julian Karoliny\IEEEauthorrefmark{1}\IEEEauthorrefmark{2}, Thomas Blazek\IEEEauthorrefmark{1}, Andreas Springer\IEEEauthorrefmark{2}, Hans-Peter Bernhard\IEEEauthorrefmark{1}\IEEEauthorrefmark{2}\\
 \IEEEauthorblockA{\IEEEauthorrefmark{1}Silicon Austria Labs GmbH, 4040 Linz~~~
 \IEEEauthorrefmark{2}Johannes Kepler University, 4040 Linz, Austria}}
\{julian.karoliny. thomas.blazek, hans-peter.bernhard\}@silicon-austria.com, andreas.springer@jku.at

 \thanks{This work is funded by the InSecTT project (https://www.insectt.eu/). InSecTT has received funding from the ECSEL Joint Undertaking (JU) under grant agreement No 876038. The JU receives support from the European Union’s Horizon 2020 research and innovation programme and Austria, Sweden, Spain, Italy, France, Portugal, Ireland, Finland, Slovenia, Poland, Netherlands, Turkey. The document reflects only the author’s view and the Commission is not responsible for any use that may be made of the information it contains.}
 }
\maketitle

\begin{abstract}
Bluetooth Low Energy (BLE) is one of the key enablers for low-power and low-cost applications in consumer electronics and the Internet of Things. The latest features such as audio and direction finding will introduce more and more devices that rely on BLE for communication. However, like many other wireless standards, BLE relies on the unlicensed 2.4\,GHz frequency band where the spectrum is already very crowded and a channel access without collisions with other devices is difficult to guarantee. For applications with high reliability requirements, it will be beneficial to actively consider channel access from other devices or standards.
In this work, we present an approach to estimate the connection parameters of multiple BLE connections outside our control and knowledge by passively listening to the channel. With this, we are able to predict future channel access of these BLE connections that can be used by other wireless networks to avoid collisions. We show the applicability of our algorithm with measurements from which we are able to identify unknown BLE connections, reconstruct their specific connection parameters, and predict their future channel access.

\end{abstract}
\vspace{0.2cm}

\section{Introduction}

Due to the Internet of Things and Industry 4.0 trends in both consumer electronics and industrial applications, an increasing number of devices are wirelessly connected. Currently, most of these devices operate in the unlicensed 2.4\,GHz \ac{ISM} band, and each new one is an additional competitor for channel access. Since reliability is a key element in wireless communication, the capability of sensing and avoiding interference becomes essential.
Popular wireless standards such as \ac{BT}, \ac{BLE}, \ac{WLAN}, and Thread include channel access methods which observe the channel before transmitting or distribute the communication over multiple channels to minimize the chance of collisions on blocked ones. Many low-power wireless sensor networks rely on deterministic channel access rather than random channel access to stay in sleep mode as long as possible. If the access to the channel is systematic, there is a good chance for other devices to identify the channel access pattern and include it in their own access scheduling. However, this systematic access might only be known to the communicating devices and may appear random to external viewers. One example is \ac{BLE}, where connected devices agree on specific transmission times for the communication. For the devices themselves, the communication happens periodically, however, due to channel hopping, the channel access might appear random from an external viewpoint. 

In this work, we propose an approach to identify active \ac{BLE} connections, estimate connection specific parameters, and predict future channel access of these connections. The current \ac{BLE} version supports two different channel access algorithms and by only passively listening to a single \ac{BLE} channel, we can reconstruct both for multiple \ac{BLE} connections in parallel. 
For the newest channel access algorithm, we will show the possibility to fully reconstruct the channel hopping pattern of a connection, including the access to channels which are not actively observed. 
The channel access information can be included in wireless networks with high reliability requirements to actively avoid collisions with \ac{BLE} connections active in the same area.
The rest of the work is organized as follows. In \cref{sec:BLE_basics} we give an overview of the important parts of the \ac{BLE} specification and how channel access is coordinated. In \cref{sec:access_prediction} we will introduce our approach for channel access prediction which is then evaluated in \cref{sec:evaluation} with measurements. Finally, conclusions are drawn in \cref{sec:conclusion}.

\subsection{Related Work}
In the unlicensed spectrum, the coexistence with other devices has to be considered. Authors in \cite{Coexistence_ISM_Band,Impacts_of_2.4-GHz_ISM_Band_Interference} studied the coexistence of different wireless communication protocols operating in the 2.4\,GHz \ac{ISM} band. In the case of \ac{BLE}, the specification \cite{bt_core_v5_2} defines no method to detect an occupied channel and reschedule the communication. However, \ac{BLE} applies adaptive frequency hopping where specific channels can be excluded for communication. Authors in \cite{Pang2021} studied the channel access mechanism of \ac{BLE} and proposed an approach to select the best channels.
Active interference mitigation is a key enabler for low power and high reliability in wireless networks. 
Specifically, it is important to enable deterministic channel access via estimation of communication slots available for interference-free communication. In particular, avoiding interference is a must if the \ac{ISM} band is used for deterministic wireless communication, as in \cite{atiq2022a} for \ac{TSN}.
Authors in \cite{Sarkar2019} proposed an approach to reconstruct \ac{BLE} connection parameters which can be directly used to predict the channel access of \ac{BLE} connections. This approach is similar to ours, however, their work focuses only on the older version of the channel access algorithms in \ac{BLE}. Due to the needed channel hopping of the sniffer in their approach, only one \ac{BLE} connection can be easily observed at a time. In our approach, only passive listening to the channel is required and both channel access algorithms of \ac{BLE} are covered. 
The prediction for \ac{BLE} advertising channels is not within the scope of this work due to its partially random nature. Authors in \cite{Braeuer2016} discussed the channel access for advertisement from a jamming perspective.


\subsection{Notation}
Scalars are written as $x$, while vectors and matrices are denoted as lower- and uppercase boldface respectively ($\mathbf{x}$ and $\mathbf{X}$). Vectors can be indexed with square brackets, e.g. $\mathbf{x}[r]$ is the $r$-th entry of the vector starting with index 0. Time indices are indicated with subscripts $x_k$. 

\section{Bluetooth Low Energy Link Layer}
\label{sec:BLE_basics}
\ac{BLE} is a \ac{wpan} technology that operates in the 2.4\,GHz \ac{ISM} band. The \ac{PHY} layer is responsible for the transmission and reception of raw data.
This work, however, targets the channel access prediction for \ac{BLE} devices. Thus, we focus on the link layer specification of the \ac{BLE} protocol. Here the channel access scheme and channel hopping are defined. 

In the \ac{BLE} link layer, the operational states of \ac{BLE} devices are defined, which can be summarized in \emph{connection state} and \emph{non-connection states}. The non-connection states include all states where no direct connection between \ac{BLE} devices is established. This includes the important \emph{advertising state} where devices announce their presence and may start setting up connections.
In the connection state, the devices exchange data in periodic connection events.
The \ac{BLE} specification \cite{bt_core_v5_2} defines 40 channels, where channels 0-36 are used for general connection events and channels 37-39 are used for advertisement.
Most of the communication in \ac{BLE} is performed in the connected state, on which we will focus in the following.

A connection between \emph{central} and \emph{peripheral} is established through an advertisement event, where the peripheral is the one that advertises its presence and the central requests a connection. In the connection process, the necessary parameters are exchanged and the two devices start communicating.
Data between central and peripheral is only exchanged during so-called connection events, which happen periodically with the connection interval $c_{\text{int}}$ (a multiple of 1.25\,ms in the range of 7.5\,ms to 4\,s). Each connection event includes at least one message sent by the central directly at the beginning. Afterwards, the peripheral and central transmit alternating if data is available. The peripheral is also allowed to skip connection events to save energy. To enumerate the connection events, we will use the connection event counter $k$, a 16-bit value that always starts at zero for the first connection event, is incremented by one for every connection event, and shall be set to zero again in case of a 16-bit number overflow (65536 to 0). 
The goal of this work is to predict the channel access of unknown \ac{BLE} connections in a certain frequency band or channel by passively listening to \ac{BLE} communication in this channel. If communication between \ac{BLE} devices would happen only on one channel, the access prediction would be trivial, as it occurs every $c_{\text{int}}$ seconds in that channel. However, \ac{BLE} applies \ac{FHSS}. A new communication channel is chosen for every new connection event, such that the access will appear random if only one channel is considered. This is also the reason why sniffing an already established \ac{BLE} connection is a challenging task, since you need to know the connection parameters to follow the channel hopping. \ac{BLE} also provides the possibility to exclude certain channels from hopping, for example if the link quality is not sufficient. The used channels are collected in ascending order in the channel map $\mathbf{c}_{\text{map}}$. \Cref{tab:con_para} summarizes the general connection parameters for \ac{BLE} connections.


\begin{table}[htb]
\centering
\caption{Connection parameters for channel hopping.}
\label{tab:con_para}
\begin{tabular}{c|l} \toprule
parameter                   & description            \\ \midrule
$c_{\text{int}}$            & connection interval    \\
$\mathbf{c}_{\text{map}}$   & list of allowed channels  \\
$n_{\text{ch}}$             & number of allowed channels \\
$k$                         & connection event counter       \\
$\text{ch}_{k}$             & calculated channel for the $k$-th event \\
$r_{k}$                     & remapping index to account for $\mathbf{c}_{\text{map}}$ \\
$\text{ch}_k'$              & used channel for the $k$-th event \\\bottomrule
\end{tabular}
\end{table}

In the \ac{BLE} specification there are currently two channel hop algorithms defined. The first one is the \ac{csa1}, which was released with the first \ac{BLE} specification, and the second is the \ac{csa2}, which was implemented with \ac{BLE} version 5.0. To tackle the whole access prediction problem, we will introduce both channel selection algorithms in the following.

\subsection{Channel Selection Algorithm \#1}
\label{sec:CSA1_Theory}
\ac{csa1} is the basic algorithm for channel selection and is used for all connections between devices where at least one device has a \ac{BLE} version below 5.0.
In addition to the parameters in \cref{tab:con_para}, the channel hopping in \ac{csa1} is defined by the hop increment $h_{\text{inc}}$. The parameters are known for the connected devices and used to calculate the communication channel for each connection event $k$.

For \ac{csa1} the unmapped channel $\text{ch}_k$ for the $k$-th connection event is calculated by
\begin{equation}
     \text{ch}_k = \text{mod} \left(\text{ch}_{k-1} + h_{\text{inc}}, 37 \right)\,, \label{math:ch_unmaped_csa1}
\end{equation}
where $\text{mod}(\cdot,\cdot)$ is the modulo operation which assures that $\text{ch}_k$ is within the allowed \ac{BLE} channels. Since \ac{BLE} also allows adaptive channel selection, we have to check whether the calculated channel $\text{ch}_k$ is in the allowed channel map $\mathbf{c}_{\text{map}}$. If $\text{ch}_k$ is an allowed channel, it is also set as mapped channel $\text{ch}_k'$ and used in the $k$-th connection event. If it is not part of the channel map, we first have to calculate the remapping index $r_k$ with
\begin{equation}
    r_k = \text{mod} \left(\text{ch}_{k}, n_{\text{ch}} \right)\,.
\end{equation}
The modulo operation ensures that $r_k$ is restricted to the number of allowed channels $ n_{\text{ch}}$. Now we can map the channel to $\text{ch}_k'$ with
\begin{equation}
  \text{ch}_k' =   \mathbf{c}_{\text{map}}\left[ r_k\right] \,.
\end{equation}
The remapping is performed only if $\text{ch}_k$ is not within the allowed channel list, otherwise no remapping is performed. For the $k$-th connection event the channel $\text{ch}_k'$ is used for communication.

An important characteristic of \ac{csa1} can be drawn from \cref{math:ch_unmaped_csa1}. It has the form of a \emph{linear congruential generator} described in \cite{Sarkar2019}. The special parameter choice for this equation in the \ac{BLE} specification makes the hop pattern repeat every 37 connection events. For example, if a connection event occurs on channel 22 for the $k=5$-th connection, it will also happen 37 connections later at $k=42$ on the same channel.

\begin{figure}[ht]
    \includegraphics[width=0.9\linewidth]{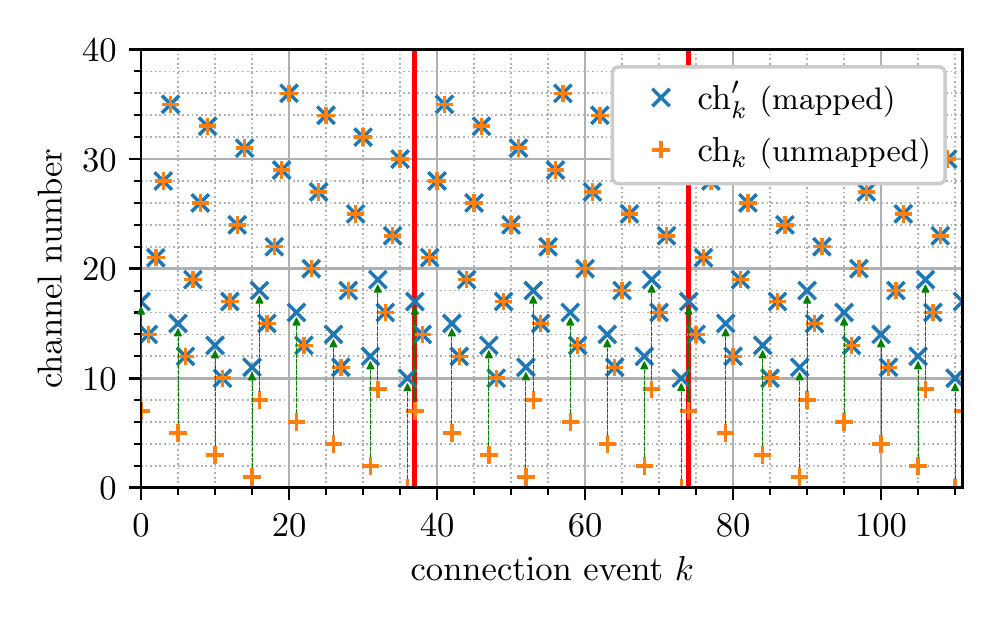} 
    \caption{Example channel hopping for \ac{csa1}.}
    \label{fig:alg1}
\end{figure}

To demonstrate this and show an example pattern of the \ac{csa1}, simulations were performed with  $c_{\text{int}} =7.5$\,ms and $h_{\text{inc}} = 7$. Additionally, adaptive channel hopping is activated where we do not allow communication in the first 10 channels.
\Cref{fig:alg1} depicts the connection event counter $k$ and the corresponding channel for this event. The red lines every 37 connection events mark the positions where the channel access pattern repeats. In this figure, we show the results for the unmapped channel $\text{ch}_k$ with orange plus-signs and the mapped channel $\text{ch}_k'$ (without using channel 0-10) with blue crosses. The remappings are indicated with green arrows. We can see for both cases that the pattern repeats exactly after 37 connections.

\subsection{Channel Selection Algorithm \#2}
In \ac{BLE} version 5.0  \ac{csa2} was added, removing some disadvantages like the short repetition interval and the unequal distribution if channels are excluded. 
Additional to the parameters in \cref{tab:con_para}, the channel identifier CI is used in \ac{csa2} to calculate the communication channel. CI is assumed to be known since it can be easily calculated with
\begin{align}
    \mathrm{CI} = \mathrm{AA}[\text{31:16}]\oplus\mathrm{AA}[\text{15:0}]\,, \label{math:CI}
\end{align}
where $\oplus$ is the bitwise xor operation and AA is the 32-bit access address that is transmitted in every \ac{BLE} packet. The notation $\mathrm{AA}[\text{15:0}]$ defines the first 16 bit of the access address and $\mathrm{AA}[\text{31:16}]$ the last, respectively.

In \ac{csa2} the channel hopping is defined by the pseudo random number $\text{prn\_e}_k$, that is calculated by the channel identifier CI and the connection event counter $k$ using the function composition
\begin{align}
    \text{prn\_e}_k= (&\oplus_{|\text{CI}} \circ f_{\text{MAM}|\text{CI}} \circ g_\text{perm} \circ f_{\text{MAM}|\text{CI}}\circ g_\text{perm} \nonumber\\ &\circ f_{\text{MAM}|\text{CI}}  \circ g_\text{perm}\circ\oplus_{|CI})(k)\,.
\label{math:prn_caluclation}
\end{align}
$\oplus_{|\text{CI}}(x)$ is a bitwise xor function conditioned on CI that can be written as 
\begin{equation}
    \oplus_{|\text{CI}}(x) = x\oplus\text{CI}\,.
\end{equation}
$g_\text{perm}$ is a permutation operation that consists of separately bit-reversing the lower and upper 8 input bits \cite{bt_core_v5_2}.\\ 
$f_{\text{MAM}|\text{CI}}(x)$ is a \ac{MAM} element, again conditioned on CI, that is defined by 
\begin{equation}
    f_{\text{MAM}|\text{CI}}(x) = \text{mod}\left(17x + \text{CI} \,,\, 2^{16} \right)\,.
\end{equation}

The unmapped channel $\text{ch}_{k}$ is now calculated using \cref{math:prn_caluclation} for every connection event $k$ as
\begin{equation}
    \text{ch}_{k} = \text{mod} \left(\text{prn\_e}_k, 37 \right) \,.\label{math:unmapped_ch_2}
\end{equation}
Compared to before, $\text{ch}_{k}$ does not depend on the previous result $k-1$, but only on current $k$ and CI.
Similar to \ac{csa1}, if $\text{ch}_k$ is not part of the channel map $\mathbf{c}_{\text{map}}$  we first have to calculate the remapping index $r_k$ using
\begin{equation}
    r_k = \left\lfloor \frac{n_{\text{ch}}\, \text{prn\_e}_k  }{  2^{16} } \right\rfloor \,, \label{math:r_k_csa2}
\end{equation}
where $\lfloor \cdot \rfloor$ is the floor function (the greatest integer less than or equal to the argument).
Now we can map the channel to $\text{ch}_k'$ with
\begin{equation}
  \text{ch}_k' =   \mathbf{c}_{\text{map}}\left[ r_k\right] \,.\label{math:mapped_ch_2}
\end{equation}
Remapping is performed only if $\text{ch}_k$ is not within the allowed channel list. For the $k$-th connection event the channel $\text{ch}_k'$ is used for the communication.

\begin{figure}[ht]
    \includegraphics[width=0.9\linewidth]{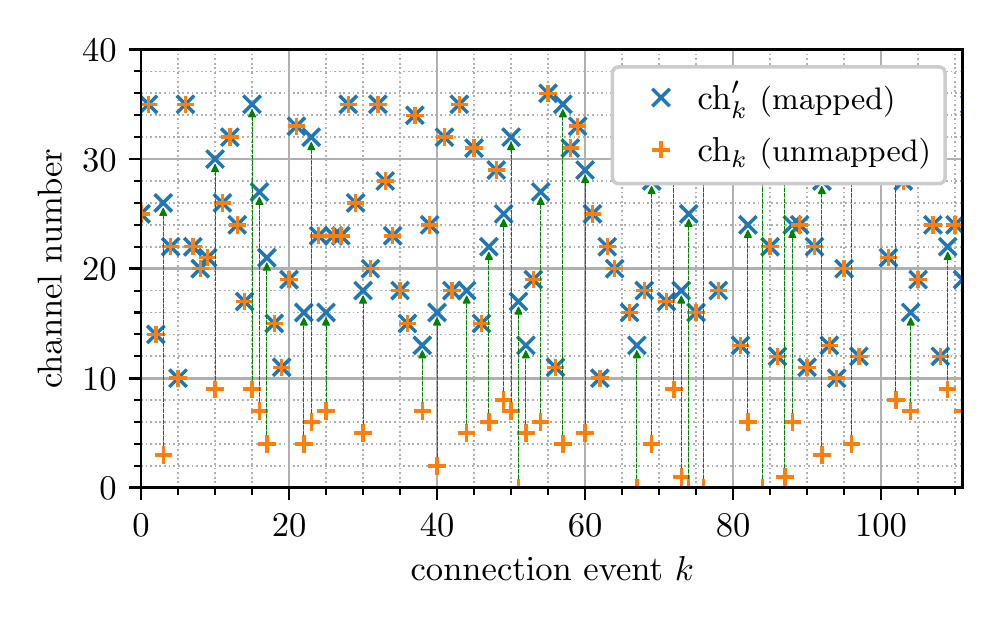} 
    \caption{Example channel hopping for \ac{csa2}.}
    \label{fig:alg2}
\end{figure}

Also for \ac{csa2} simulations were performed with $c_\text{int}=7.5\,\text{ms}$ and a $\mathbf{c}_{\text{map}}$ that excludes the first 10 channels. As access address we use \texttt{0xB0A1CD9D} to calculate CI by \cref{math:CI} and start for $k=0$. \Cref{fig:alg2} depicts the connection event counter $k$ and the corresponding channel for this event. The results for the unmapped channel $\text{ch}_k$ are marked with orange plus-signs and the mapped channel $\text{ch}_k'$ with blue crosses. Again, the remapping is indicated with green arrows.
Since the repetition interval of the pattern is much larger (65536 instead of 37 connection events), no repetition is visible here. Compared to \ac{csa1}, the remapping in \cref{fig:alg2} is not always to the same channel, it is uniformly distributed across all $n_{\text{ch}}$   channels \cite{Pang2021}. 

\section{Channel Access Prediction}
\label{sec:access_prediction}
In this section, we present an approach to predict the channel access of \ac{BLE} devices for both channel selection algorithms.
Current approaches in literature require additional channel hopping of the sniffer, which restricts the evaluation to only one \ac{BLE} connection at a time. With our approach, we are able to evaluate multiple \ac{BLE} connections in parallel. 

If one \ac{BLE} channel is sniffed, packets of multiple connections can be observed. However, these packets can be easily separated by the access address which is transmitted at the start of each packet \cite{bt_core_v5_2}. The header of a \ac{BLE} payload is not encrypted \cite{Caesar2022}, which allows us to explicitly filter the central node messages that are transmitted at the beginning of all connection events. 
We define the channel we are passively listening to as $\text{ch}^\text{sniff}$. In the following, the reconstruction is only described for a single device of a \ac{BLE} connection (e.g. central), however, it works also for multiple devices in parallel since they can be distinguished by the access address and header.
By listening passively to $\text{ch}^\text{sniff}$, the only measurement that is available is the time $\mathbf{t}_a \in \mathbb{R}^{N_a}$ of the message reception, where $N_a$ is the number of measurements we collected as
\begin{equation}
    \mathbf{t}_a  = \left[t_{0},\,t_{1},\,t_{2},\,\dots,\,t_{{{N_a}-1}}  \right]\,.
\end{equation}
One important characteristic of $\mathbf{t}_a$ is that between measurements there is always an integer multiple of the connection interval, which can be written as 
\begin{equation}
    \mathbf{\Delta t}_a = \big[ \underbrace{t_{1}-t_{0}}_{x_1 c_\text{int}} ,\,\underbrace{t_{2}-t_{1}}_{x_2 c_\text{int}},\,\underbrace{t_{3}-t_{2}}_{x_3 c_\text{int}},\,\dots  \big]\equiv \mathbf{x}\,c_{\text{int}}\,, \label{math:delta_t}
\end{equation}
where $x_i$ is an integer number that may vary for each entry and $c_\text{int}$ is the connection interval that we need to estimate. In our approach we use $\mathbf{t}_a$ to reconstruct all needed parameters for the channel access prediction.
 
However, we first need to reconstruct the connection interval $c_\text{int}$ and determine which of the two channel selection algorithms is used. For this we have to distinguish three different cases. First, if all entries in \cref{math:delta_t} are the same, \ac{csa1} was used and we can estimate $c_\text{int}$ by simply dividing the entries by 37, which is the repetition interval. In the second case, the entries are different, but they repeat after a few measurements. This is due to the short repetition interval of \ac{csa1} which can be seen in \cref{fig:alg1} by observing channel 10. Here we have two observations per repetition interval because of the remapping from channel 0. In this example we would measure $\mathbf{\Delta t}_a = \left[25 c_\text{int},\,12 c_\text{int},\,25 c_\text{int},\,12 c_\text{int},\,\dots   \right]$, were we clearly see the characteristic of \ac{csa1}. Since there is always an integer number of connection events between the measurements, $c_\text{int}$ can be estimated by 
\begin{align}
    \hat{c}_\text{int} = \text{GCD}(\mathbf{\Delta t}_a)\,, \label{math:gcd}
\end{align}
where GCD is a function that calculates the greatest common divider.
In the third case, \cref{math:delta_t} does not show any repetition, so \ac{csa2} was used. Here, $c_\text{int}$ can also be calculated with \cref{math:gcd}.
To account for measurement errors, a rounding to 1.25\,ms steps can be applied, which is the resolution of $c_\text{int}$ defined in the \ac{BLE} specification \cite{bt_core_v5_2}.
For both algorithms, we give now an approach to predict the future channel access.

\subsection{Channel Selection Algorithm \#1}
\label{sec:CSA1_Prediction}
As mentioned in \cref{sec:CSA1_Theory}, the channel access pattern repeats every 37 connection events. For example, if we listen to $\text{ch}^\text{sniff}$ and observe packets from a device at the connection event $k=2$ and $k=4$,  the channel access will also be observable for $k=39$ and $k=41$. As a result, for \ac{csa1} we only need the connection interval $c_\text{int}$ and measure for a period of $37\,c_\text{int}$. Every access from the device to the sniffed channel within this period will appear again after 37 connection events, the starting point is not important.

With $c_\text{int}$  and the observations per repetition, the future channel access can be predicted by simply adding $37\,c_\text{int}$ to the current observation. To account for measurement errors and clock drifts, a Kalman filter \cite{welch1995introduction} with a constant velocity motion model can be used to stay synchronized with the connection interval.
By listening only passively to one channel, it is not possible to estimate the channel map $\mathbf{c}_{\text{map}}$  and the hop increment $h_{\text{inc}}$. If $\text{ch}^\text{sniff}$ is changed during the sniffing procedure, it is possible to estimate $h_{\text{inc}}$ and $\mathbf{c}_{\text{map}}$ as described in \cite{Sarkar2019}. However, since our use case is the prediction of channel access for one channel, these parameters are not needed.
\subsection{Channel Selection Algorithm \#2}
Since the \ac{csa2} has a more complex structure and lacks the short repetition interval, all parameters listed in \cref{tab:con_para} have to be estimated for the access prediction.
For the estimation, we propose a two-step approach in which we first reconstruct the connection event counter $k$ and then the channel map $\mathbf{c}_{\text{map}}$.
\subsubsection{Reconstruct the connection event counter}
With $c_\text{int}$ computed by \cref{math:gcd}, we know how many channel hops happened between the measured connection events in $\mathbf{t}_a$. The sniffed \ac{BLE} connection has most likely not just started, thus the connection event counter $k$ will be some number between 0 and 65535. We define $k_\text{init}$ as the connection event corresponding to the first observation of the \ac{BLE} connection. i.e. the value of $k$ for $\mathbf{t}_a[0]$. With $k_\text{init}$ we can determine the value of $k$ for all measurements in $\mathbf{t}_a$ and also for all future measurements. For this, we construct the binary vector $\textbf{c}^\text{meas} \in \{0,1\}^{N_{m}}$ using $\mathbf{\Delta t}_a$ as
\begin{equation}
    \mathbf {c}^{\text{meas}} = \left[\mathbf I_{1\times x_1},\mathbf I_{1\times x_2},\dots, \mathbf I_{1\times x_{N_a-1}} \right], 
\end{equation}
where $x_i$ can be calculated with $c_{\text{int}}$ from \cref{math:delta_t} and  $\mathbf I_{M\times N}$ is a $M\times N$ matrix where the entries are given as 
\begin{equation}
    I_{n, m} = \left\{
    \begin{array}{rl}
      1   & m=n  \\
        0  & \text{otherwise} 
    \end{array}\right.\,.
\end{equation}
$\mathbf {c}^{\text{meas}}$ basically lists all $N_m$ connection events that happen during the measurement, where the entries are 1 if there was a observation in $\text{ch}^\text{sniff}$ and 0 otherwise. Additionally, we construct the binary vector $\textbf{c}^\text{ref} \in \{0,1\}^{65536}$ for all possible $k \in [0,\,65535]$
\begin{align}
    \textbf{c}^\text{ref}[k] = 
    \begin{cases}
        1 ,& \text{if } \text{ch}_k =\text{ch}^\text{sniff}\\
        0,              & \text{otherwise} 
\end{cases}\,, \label{math:c_ref}
\end{align}
using \cref{math:prn_caluclation,math:unmapped_ch_2} to calculate all possible $\text{ch}_k$. Here, the unmapped channel number is used, since we have no prior knowledge of $\mathbf{c}_{\text{map}}$.
The idea now is to find the position where $\textbf{c}^\text{ref}$ and $\textbf{c}^\text{meas}$ have the highest correlation. For this, we calculate the circular cross-correlation between both and estimate the maximum as 
\begin{align}
	r[k]=&\sum_{m=0}^{{N_{m}-1}} \mathbf{c}^\text{ref}\left[\text{mod}\left(m+k, 2^{16}\right)\right] \mathbf{c}^\text{meas}[m]\,,\\
   k_\text{init} =& \arg\max_k\big(r[k]\big)\,.
\end{align}
However, to predict all channel accesses of a \ac{BLE} connection, we also need to consider the remapping in \cref{math:mapped_ch_2,math:r_k_csa2}. Therefore, $\mathbf{c}_{\text{map}}$ also needs to be estimated.
%
\subsubsection{Estimate the channel map}
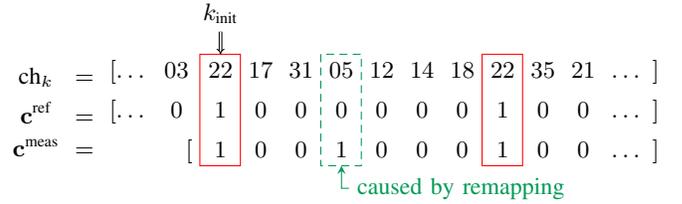
\begin{figure}[ht]
\begin{tikzpicture}
        \small
       \matrix (m)[
        matrix of math nodes,
        nodes in empty cells
        ] {
        \text{ch}_k & =  \\
        \mathbf{c}^\text{ref}& =   \\
        \mathbf{c}^\text{meas} & =  \\
        } ;
        
    \begin{scope}[xshift=4.4cm]
        
        \matrix (m)[
        matrix of math nodes,
        nodes in empty cells] {
        \small[\dots &03& 22 & 17 & 31 & 05 & 12 & 14 & 18 & 22 & 35 & 21 &   \dots~\small] \\
        \small[\dots &0&  1 & 0  & 0  & 0  & 0  & 0  & 0  & 1  & 0  & 0  &  \dots~\small] \\
        &~~~\small[& 1     & 0  & 0  & 1  & 0  & 0  & 0  & 1  & 0  & 0  &  \dots~\small] \\
        } ;
        \draw[color=red] (m-1-3.north west) rectangle ([shift={(0.07,0)}]m-3-3.south east);
        \draw[color=red] (m-1-10.north west) rectangle ([shift={(0.07,0)}]m-3-10.south east);
        \draw[color=ForestGreen,densely dashed] (m-1-6.north west) rectangle ([shift={(0.07,0)}]m-3-6.south east);
        \draw[color=black,double,implies-](m-1-3.north) -- +(0,0.3)node[above] {$k_\text{init}$};
         \draw[color=ForestGreen,stealth-](m-3-6.south) |- +(0.1,-0.3)node[right] {caused by remapping};
    \end{scope}
\end{tikzpicture}
\caption{Process to reconstruct the channel map $\mathbf{c}_{\text{map}}$.}
\label{fig:cmap_est}
\end{figure}

One advantage of \ac{csa2} is that, compared to \ac{csa1}, the remapping of a certain channel appears not always to the same other channel (compare the remapping in \cref{fig:alg1} and \cref{fig:alg2}). Due to this, an unused channel will always be mapped to the sniffed channel $\text{ch}^\text{sniff}$ if we observe the channel long enough. This characteristic can be used to completely reconstruct the channel map without listening to all channels. The schematic of this approach is shown in \cref{fig:cmap_est} where we assume that we are passively listening at $\text{ch}^\text{sniff} = 22$. Here we again use $\mathbf{c}^\text{ref}$ for all possible $k$ and compare it with $\mathbf{c}^\text{meas}$. The two vectors are now aligned at $k_\text{init}$, as highlighted in the figure. For all entries where the unmapped channel $\text{ch}_k$ is $\text{ch}^\text{sniff}$, both $\mathbf{c}^\text{ref}$ and $\mathbf{c}^\text{meas}$ have 1 as entry. In the cases where the calculated channel is not the sniffed one, we have no observations in general. However, in some cases we will have an observation in $\mathbf{c}^\text{meas}$ but not in $\mathbf{c}^\text{ref}$. This is caused by remapping, for example in \cref{fig:cmap_est} for channel 5 highlighted with the green dashed rectangle. In this case, we can be sure that the corresponding channel, e.g. here channel 5, is not part of the channel map. However, the counter-argument cannot be used. If we have no measurement we do not know if the channel access happens at the planed channel or was remapped to a channel we are not observing. With this approach, $\mathbf{c}_{\text{map}}$ and $n_{\text{ch}}$  can be  reconstructed iteratively.

To compute the expected number of required measurements for reconstructing  \ac{csa2}, we assume an uniform distribution both for the mapping and remapping \cite{Pang2021}. Thus,  we expect an excluded channel to be remapped to the sniffed channel on average on the $n_{\text{ch}}$th access. The \emph{coupon collector problem} \cite{FLAJOLET1992207}
gives us an estimate for the expected number of channel access needed to access every remapped channel at least once. Combined with the probability of accessing an excluded channel $p_{rem}=\tfrac{37-n_{\text{ch}}}{37}$, we estimate the  number of channel hops to be measured to be $N_m =  n_{\text{ch}}(37-n_{\text{ch}})\left(\sum_{i=1}^{37-n_{\text{ch}}}\tfrac{1}{i}\right)p_{rem}^{-1}$, with the worst case occurring for $n_{ch}=28$, resulting in  $N_m \approx 2932$, or 21.99\,s for $c_\text{int}=7.5$\,ms.

With CI, $k$, $\mathbf{c}_{\text{map}}$, and $n_{\text{ch}}$ we have now all parameter to perform the same hop calculation as the \ac{BLE} devices of the sniffed communication. As a result, we are able to predict the access to all used \ac{BLE} channels and not only the sniffed one. This is unique for this algorithm, since for \ac{csa1} we would need to sniff multiple channels for this. Similarly to \ac{csa1}, it is necessary to account for clock drifts and stay synchronized with the connection interval for the prediction step.

\section{Measurement Results}
\label{sec:evaluation}
With our approach the channel hopping can be predicted exactly and prediction problems can only occur due to measurement errors. Therefore, instead of simulations, we directly show the applicability of our approach with measurements.
Our measurement setup consists of six Nordic NRF52840 \ac{BLE} devices that form three \ac{BLE} connection pairs and one sniffer based on the \emph{Ubertooth One}. The \ac{BLE} devices are running the \emph{heart-rate monitor} sample of the Zephyr Project \cite{ZephyrGit}, modified to allow configuring $c_{\text{int}}$ and  $\mathbf{c}_{\text{map}}$. The \emph{Ubertooth One} demodulates the raw signals of one \ac{BLE} channel and provides the measured bitstream. With this we can measure multiple \ac{BLE} connections in parallel and separate individual ones by their access address.
The measurements conducted during this work are published as open-source in \cite{ble_dataset}, where also a more detailed description of the setup is provided.
For the following evaluation, the measurement set \emph{dataset\_ubertooth/set\_1} in \cite{ble_dataset} was used. The configuration of the \ac{BLE} devices is listed in \cref{tab:meas_parameter}.

\begin{table}[htb]
\centering
\caption{Configuration parameters and prediction results.}
\label{tab:meas_parameter}
\begin{tabular}{ccc|cc} \toprule
\multicolumn{1}{c}{CSA} & \multicolumn{1}{c}{$c_{\text{int}}$ [ms]} & \multicolumn{1}{c}{$\mathbf{c}_{\text{map}}$} & \multicolumn{1}{|c}{$\hat{c}_{\text{int}}$ [ms]}& \multicolumn{1}{c}{RSME [$mu$s]} \\\midrule
1 &  18.75  &   \texttt{0x1FFFFFFC00}   &18.747& 0.1163  \\
2 &  12.50  &   \texttt{0x1E00E00700}   &12.498& 0.0699  \\
2 &  ~7.50   &   \texttt{0x1FFFFFFC00}  &~7.499& 0.1187  \\\bottomrule
\end{tabular}
\end{table}
For two \ac{BLE} connection pairs \ac{csa2} and for one the older \ac{csa1} was used. The devices communicated with three different values of $c_{\text{int}}$ and we applied two different values of $\mathbf{c}_{\text{map}}$. \Cref{tab:meas_parameter} provides the hexadecimal representation of the channel maps, where \texttt{0x1FFFFFFC00} corresponds to a channel map where we do not use the first 10 \ac{BLE} channels (as in \cref{fig:alg1,fig:alg2}) and \texttt{0x1E00E00700} uses channels between the widely used \ac{WLAN} channels 1, 6 and 11. For the sniffer, we choose a center frequency of 2.45\,GHz, which corresponds to \ac{BLE} channel 22.

\begin{figure}[ht]
    \includegraphics[width=0.90\linewidth]{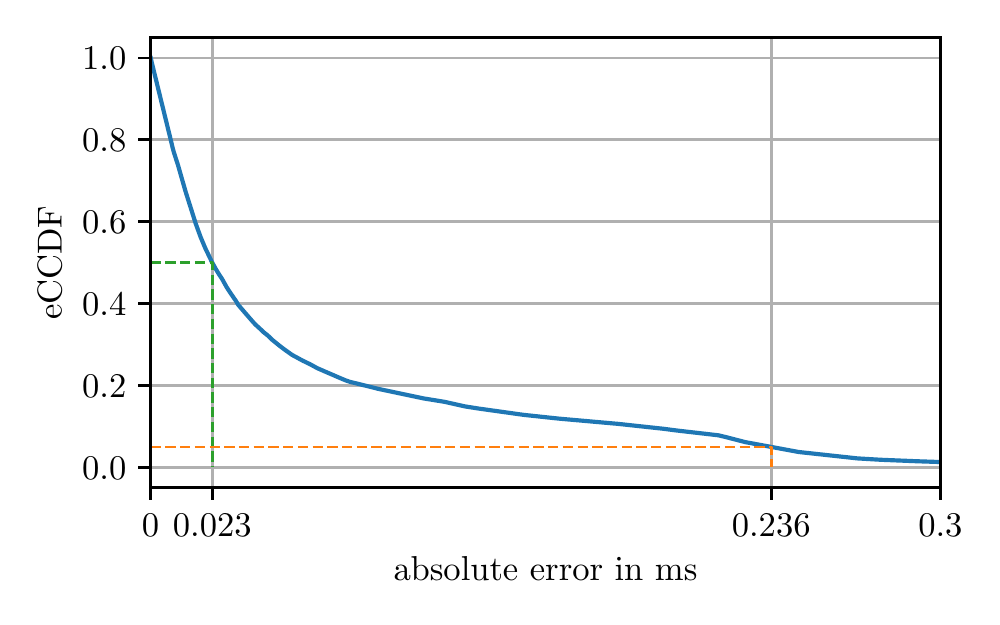} 
    \caption{\ac{eccdf} of the absolute error between all combined measurements and the corresponding predictions.}
    \label{fig:ecdf_abs_error}
\end{figure}

The measurement duration was 400\,s, where we used the first 100\,s to estimate all needed parameters as described in \cref{sec:access_prediction} and performed predictions on the remaining 300\,s.
For the prediction, we first evaluated the connection counter $k$ of the latest measurement and then calculated the future connection events for the observed channel using the method described in \cref{sec:access_prediction}. To predict the access time a few connection events ahead, a simple multiplication with $\hat{c}_{\text{int}}$ is sufficient to accurately predict the channel access. However, to account for clock drift and measurement errors, we added a Kalman filter with a constant velocity motion model to stay synchronized and predict the channel access time with a higher accuracy.
\Cref{tab:meas_parameter} lists the \ac{RMSE} between the measured $\mathbf{t}_a$ and the predicted one for the remaining 300\,s, and also the estimated $\hat{c}_{\text{int}}$.
The estimated connection intervals match the configured but are slightly lower due to clock differences between sniffer and \ac{BLE} devices. 
For all active \ac{BLE} connections, we were able to reconstruct the hopping pattern and stay synchronized to the individual connections. We could achieve an average \ac{RMSE} of 0.1016\,ms for the access time prediction.
\cref{fig:ecdf_abs_error} shows the \ac{eccdf} of the absolute error between the estimated channel access and the measured one, where we additionally marked the 5\% and 50\% probability including the corresponding absolute error. We can see that 50\% of the estimation shows an error below 0.023\,ms, while only for 5\% of the measurements the error exceeded 0.236\,ms. With this accuracy, it is easily possible to stay synchronized and continuously perform predictions of the channel access. Since the channel access is deterministic and as soon as the corresponding parameters are reconstructed, the channel access can be theoretically estimated exactly. The presented error is due to the measurement accuracy and missing measurements.

As long as the parameters of the observed connections do not change, e.g. an update of $c_{\text{int}}$ or $\mathbf{c}_{\text{map}}$, we are able to calculate the channel hopping in a similar way as the \ac{BLE} devices and continue predicting the channel access. To account for changes in the parameters, the corresponding steps in \cref{sec:access_prediction} have to be repeated. However, since the procedure is iterative, it is beneficial to perform the parameter estimation continuously for new measurements. This allows to detect changes and immediately update the algorithm.

\section{Conclusion}
\label{sec:conclusion}
In order to have a more deterministic channel access in \ac{ISM} bands, we presented an approach to predict the channel access of multiple unknown \ac{BLE} connections. We are able to identify active \ac{BLE} connections and reconstruct their relevant connection parameters. Based on the  standardized access schemes for \ac{BLE} we are thus able to predict future channel access for multiple devices in parallel.
These predictions can be used by wireless networks with high reliability requirements operating in close proximity to the \ac{BLE} devices to reschedule their own communication and avoid using time slots and channels with predicted interference. 
For the latest channel selection algorithm in \ac{BLE} version 5.0 and above, our algorithm allows to completely reconstruct the connection parameters. This gives us the unique possibility to predict the future channel access for all used \ac{BLE} channels while only listening passively to a single one. The applicability of this approach is demonstrated  by measurements and identification of three \ac{BLE} links including their channel access.



\bibliographystyle{IEEEtran}
\bibliography{refs}

\end{document}